\documentclass[twocolumn,showpacs,preprintnumbers,prl]{revtex4}
\usepackage{epsfig,amsmath,amssymb}

\newcommand{\beq}{\begin{equation}}
\newcommand{\eeq}{\end{equation}}
\newcommand{\bea}{\begin{eqnarray}}
\newcommand{\eea}{\end{eqnarray}}
\newcommand{\Fig}[1]{Fig.\,\ref{#1}}

\newcommand{\Eq}[1]{Eq.\,(\ref{#1})}
\newcommand{\eq}[1]{(\ref{#1})}
\newcommand{\Eqs}[1]{Eqs.\,(\ref{#1})}
\newcommand{\Eqsand}[2]{Eqs.\,(\ref{#1}) and (\ref{#2})}
\newcommand{\Eqsto}[2]{Eqs.\,(\ref{#1})--(\ref{#2})}

\newcommand{\Tab}[1]{Table\,\ref{#1}}
\newcommand{\f}{\frac}
\newcommand{\non}{\nonumber}

\newcommand{\as}{\alpha_s}
\newcommand{\aem}{\alpha}
\newcommand{\aemMSbar}{\alpha_{\overline{\rm MS}}}
\newcommand{\sws}{\sin^2 \theta_{\scriptscriptstyle W}}
\newcommand{\swsMSbar}{\sin^2 \hat{\theta}_{\scriptscriptstyle
W}^{\overline{\rm MS}}}    
\newcommand{\swq}{\sin^4 \theta_{\scriptscriptstyle W}}
\newcommand{\MW}{M_{\scriptscriptstyle W}}
\newcommand{\MZ}{M_{\scriptscriptstyle Z}}
\newcommand{\mc}{m_c}
\newcommand{\mb}{m_b}
\newcommand{\mt}{m_t}
\newcommand{\mtpole}{M_t}

\newcommand{\muh}{\mu_{\scriptscriptstyle W}}

\newcommand{\mut}{\mu_t}
\newcommand{\muc}{\mu_c}
\newcommand{\mub}{\mu_b}

\newcommand{\BR}{{\cal{B}}}
\newcommand{\GeV}{{\rm \ GeV}}
\newcommand{\TeV}{{\rm \ TeV}}
\newcommand{\MSbar}{\overline{\rm MS}}
\newcommand{\re}{{\rm Re}}
\newcommand{\im}{{\rm Im}}
\newcommand{\ord}{{\cal O}}

\newcommand{\Ktopinunu}{K^+ \to \pi^+ \nu \bar{\nu}}

\newcommand{\KLtopinunu}{K_L \to \pi^0 \nu \bar{\nu}}

\newcommand{\Ktopienu}{K^+ \to \pi^0 e^+ \nu}
\newcommand{\Ktopinunus}{K \to \pi \nu \bar{\nu}}

\newcommand{\Pc}{P_c}
\newcommand{\dPuc}{\delta P_{c, u}}

\begin{document}

%\allowdisplaybreaks

\preprint{TUM-HEP-599-05; IPPP/05/50; DCPT/05/100;
FERMILAB-Pub-05-347-T; arXiv:hep-ph/0508165} 

\title{
\boldmath
Rare Decay $\Ktopinunu$ at the Next-to-Next-to-Leading Order in QCD   
\unboldmath}
\author{Andrzej J.~Buras,${}^1$ Martin Gorbahn,${}^2$
Ulrich Haisch,${}^3$ and Ulrich Nierste${}^3$}   

\affiliation{
$^1\!\!\!$ Physik Department, Technische Universit\"at M\"unchen,
D-85748 Garching, Germany \\
$^2\!\!\!$ IPPP, Physics Department, University of Durham, DH1 3LE,
Durham, UK \\  
$^3\!\!\!$ Theoretical Physics Department, Fermilab, Batavia, IL 60510,
USA 
}

\date{December 16, 2005}

\begin{abstract}
\noindent
We calculate the charm quark contribution to the rare decay
$\Ktopinunu$ in the next-to-next-to-leading order of QCD. This new
contribution reduces the theoretical uncertainty in the relevant
parameter $\Pc$ from $\pm 10.1 \%$ down to $\pm 2.4 \%$, corresponding
to scale uncertainties of $\pm 1.3 \%$, $\pm 1.0 \%$, $\pm 0.006$ and
$\pm 1.2^\circ$ in $\BR (\Ktopinunu)$ and in $| V_{td} |$, $\sin 2  
\beta$ and $\gamma$ extracted from the $\Ktopinunus$ system. The error
in $\Pc = 0.37 \pm 0.04$ is now fully dominated by the current
uncertainty of $\pm 3.8 \%$ in the charm quark mass $\mc$. We
find $\BR (\Ktopinunu) = \left ( 8.0 \pm 1.1 \right ) \times
10^{-11}$, where the quoted error stems almost entirely from the
present uncertainties in $\mc$ and the Cabibbo-Kobayashi-Maskawa
elements.  
\end{abstract}
\maketitle

The rare process $\Ktopinunu$ belongs to the theoretically cleanest
decays in the field of $K$ and $B$ mesons. As it offers in
conjunction with $\KLtopinunu$ a very clean determination of the
standard unitarity triangle \cite{UT}, a comparison of the information
obtained from the $\Ktopinunus$ system with the one from $B$ decays
provides a critical and truly unique test of the
Cabibbo-Kobayashi-Maskawa (CKM) mechanism in the standard model (SM) 
\cite{Gino, Buras:2004uu}. Even if these $K$- and $B$-physics
predictions agree, $\Ktopinunu$ will allow to discriminate between
different extensions of the SM \cite{Gino, Buras:2004uu}, by probing
effective scales of new physics operators of up to a several $\!\!
\TeV$ or even higher \cite{D'Ambrosio:2002ex}.       

In the SM the decay $\Ktopinunu$ proceeds through $Z$-penguin and
electroweak box diagrams, which are sensitive to short-distance
dynamics. As the required hadronic matrix elements can be extracted,
including isospin breaking corrections \cite{Marciano:1996wy}, from
the accurately measured leading semileptonic decay $\Ktopienu$, and
the remaining long-distance contributions turn out to be small
\cite{Isidori:2005xm}, and in principle calculable by means of lattice
QCD \cite{Isidori:2005tv}, theoretical computations of the relevant
decay rate can reach an exceptionally high degree of precision.      

After summation over the three neutrino flavors the resulting
branching ratio for $\Ktopinunu$ can be written as
\cite{Buras:2004uu, Isidori:2005xm, Buchalla:1993wq, Buchalla:1998ba}
\begin{widetext}
\begin{gather} 
\label{eq:BR}
\BR ( \Ktopinunu) = \kappa_+ \left [ \left ( \f{\im
\lambda_t}{\lambda^5} X(x_t) \right )^2 + \left ( \f{\re 
\lambda_t}{\lambda^5} X(x_t) + \f{\re \lambda_c}{\lambda} \left (\Pc +
\dPuc \right ) \right )^2 \right ] \, , \\ \label{eq:kappap}
\kappa_+ \equiv r_{K^+} \f{3 \aem^2 \, \BR (\Ktopienu)}{2 \pi^2 \swq} 
\lambda^8 = \left ( 5.04 \pm 0.17 \right ) \times 10^{-11} \left (
\f{\lambda}{0.2248} \right )^8 \, . 
\end{gather}
\end{widetext}
Here $\lambda_i \equiv V^\ast_{is} V_{id}$ denote the relevant CKM 
factors, while $\dPuc = 0.04 \pm 0.02$ encodes the long-distance
contributions calculated recently in \cite{Isidori:2005xm}, and the
parameter $r_{K^+} = 0.901 \pm 0.027$ summarizes isospin breaking
corrections in relating $\Ktopinunu$ to $\Ktopienu$
\cite{Marciano:1996wy}. The apparent strong dependence of $\BR
(\Ktopinunu)$ on $\lambda \equiv | V_{us} |$ is spurious as $\Pc$ and
$\dPuc$ are proportional to $1/\lambda^4$. In quoting the value for
$\Pc$ and $\BR (\Ktopinunu)$ we will set $\lambda = 0.2248$
\cite{Charles:2004jd}. The electromagnetic coupling $\aem$ and the
weak mixing angle $\sws$ entering $\BR (\Ktopinunu)$ are naturally
evaluated at the electroweak scale \cite{Bobeth:2003at}. Then the
leading term in the heavy top expansion of the electroweak two-loop
corrections to $X (x_t)$ amounts to typically $-1 \%$ for the modified
minimal substraction scheme ($\MSbar$) definition of $\aem$ and $\sws$
\cite{Buchalla:1997kz}. In obtaining the numerical value of
\Eq{eq:kappap} we have employed $\aem \equiv \aemMSbar (\MZ) =
1/127.9$, $\sws \equiv \swsMSbar = 0.231$, and $\BR (\Ktopienu) = \left
( 4.93 \pm 0.07 \right ) \times 10^{-2}$ \cite{PDG}.  

The function $X(x_t)$ in \Eq{eq:BR} depends on the top quark $\MSbar$
mass through $x_t \equiv \mt^2 (\mut)/\MW^2$. It originates from
$Z$-penguin and electroweak box diagrams with an internal top
quark. As the relevant operator has a vanishing anomalous dimension
and the energy scales involved are of the order of the electroweak
scale or higher, the function $X(x_t)$ can be calculated within
ordinary perturbation theory. It is known through next-to-leading
order (NLO) \cite{Buchalla:1998ba, X}, with a scale uncertainty due to
the top quark matching scale $\mut = \ord (\mt)$ of only $\pm 1
\%$. Converting the top quark pole mass of $\mtpole = \left ( 172.7
\pm 2.9 \right ) \!\! \GeV$ \cite{Group:2005cc} at three loops to $\mt
(\mtpole)$ \cite{Melnikov:2000qh} and relating $\mt (\mtpole)$ to $\mt
(\mt) = \left ( 163.0 \pm 2.8 \right ) \!\! \GeV$ using the one-loop
renormalization group (RG), we find $X (x_t) = 1.464 \pm 0.041$. The
given uncertainty combines linearly an error of $\pm 0.028$ due to the
error of $\mt (\mt)$ and an error of $\pm 0.013$ obtained by varying
$\mut$ in the range $60 \GeV \le \mut \le 240 \GeV$.                  

The calculable parameter $\Pc$ entering \Eq{eq:BR} results from
$Z$-penguin and electroweak box diagrams involving internal charm
quark exchange. As now both high- and low-energy scales, namely, $\muh
= \ord (\MW)$ and $\muc = \ord (\mc)$, are involved, a complete RG
analysis of this term is required. In this manner, large logarithms
$\ln (\muh^2/\muc^2)$ are resummed to all orders in $\as$. At the
leading order such an analysis has been performed in \cite{LO}. The
large scale uncertainty due to $\muc$ of $\pm 26 \%$ in this result
was a strong motivation for the NLO analysis of this contribution
\cite{Buchalla:1993wq, Buchalla:1998ba}. 

\begin{table}[!t]
\caption{Input parameters used in the numerical analysis of $\Pc$,
$\BR (\Ktopinunu)$, $| V_{td} |$, $\sin 2 \beta$ and $\gamma$.}  
\begin{tabular}{l@{\hspace{5mm}}c@{\hspace{5mm}}c}
\hline \hline
Parameter & Value $\pm$ Error & Reference \\[0.25mm]
\hline
&&\\[-3.5mm]
$\mc (\mc)$ \hspace{0.2mm} [GeV] & $1.30 \pm 0.05$ & \cite{charm}, our
average \\     
$\as (\MZ)$ & $0.1187 \pm 0.0020$ & \cite{PDG} \\[0.5mm]
$\im \lambda_t$ \hspace{0.2mm} $[10^{-4}]$ & $1.407^{+0.096}_{-0.098}$
& \cite{Charles:2004jd} \\[0.5mm] 
$\re \lambda_t$ \hspace{0.25mm} $[10^{-4}]$ & $-3.13^{+0.20}_{-0.17}$ 
& \cite{Charles:2004jd} \\[0.5mm] 
$\re \lambda_c$ & $-0.22006^{+0.00093}_{-0.00091}$ &
\cite{Charles:2004jd} \\[1mm]   
\hline \hline
\end{tabular}
\label{tab:input}
\end{table}

Performing the RG running from $\muh$ down to $\mub = \ord (\mb)$ in
an effective five-flavor theory and the subsequent evolution from
$\mub$ down to $\muc$ in an effective four-flavor theory, we obtain at
the NLO          
\beq \label{eq:PcNLO}
\begin{split}
\Pc & = 0.367 \pm 0.037_{\rm theor} \pm 0.033_{\mc} \pm
0.009_{\as} \\
& = \left ( 0.37 \pm 0.06 \right ) \left ( \f{0.2248}{\lambda} \right
)^4 \, ,  
\end{split}
\eeq 
where the parametric errors correspond to the ranges of the
charm quark $\MSbar$ mass $\mc (\mc)$ and the strong coupling constant
$\as (\MZ)$ given in \Tab{tab:input}. We note that the final error has
only an illustrative character, since the partial uncertainties are
not statistically distributed. Numerically, it is the mean of the
values obtained by adding the individual errors once linearly and once
in quadrature. The same way of combining errors will be applied in
\Eqs{eq:BRNLO}, \eq{eq:PcNNLO}, and \eq{eq:BRNNLO}.              

\begin{figure}[!t]
\begin{center}
\scalebox{0.7}{% GNUPLOT: LaTeX picture with Postscript
\begingroup%
  \makeatletter%
  \newcommand{\GNUPLOTspecial}{%
    \@sanitize\catcode`\%=14\relax\special}%
  \setlength{\unitlength}{0.1bp}%
\begin{picture}(3600,2160)(0,0)%
\special{psfile=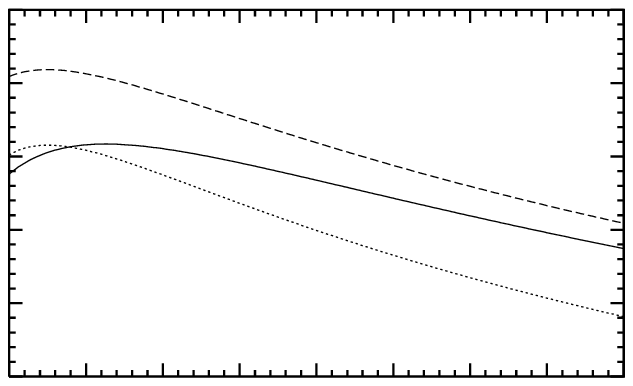 llx=0 lly=0 urx=360 ury=216 rwi=3600}
\put(1975,50){\makebox(0,0){\scalebox{1.3}{$\muc$ [GeV]}}}%
\put(100,1180){%
\special{ps: gsave currentpoint currentpoint translate
270 rotate neg exch neg exch translate}%
\makebox(0,0)[b]{\shortstack{\scalebox{1.3}{$\Pc$}}}%
\special{ps: currentpoint grestore moveto}%
}%
\put(3450,200){\makebox(0,0){ 3}}%
\put(3081,200){\makebox(0,0){ 2.75}}%
\put(2713,200){\makebox(0,0){ 2.5}}%
\put(2344,200){\makebox(0,0){ 2.25}}%
\put(1975,200){\makebox(0,0){ 2}}%
\put(1606,200){\makebox(0,0){ 1.75}}%
\put(1238,200){\makebox(0,0){ 1.5}}%
\put(869,200){\makebox(0,0){ 1.25}}%
\put(500,200){\makebox(0,0){ 1}}%
\put(450,2060){\makebox(0,0)[r]{ 0.42}}%
\put(450,1708){\makebox(0,0)[r]{ 0.4}}%
\put(450,1356){\makebox(0,0)[r]{ 0.38}}%
\put(450,1004){\makebox(0,0)[r]{ 0.36}}%
\put(450,652){\makebox(0,0)[r]{ 0.34}}%
\put(450,300){\makebox(0,0)[r]{ 0.32}}%
\end{picture}%
\endgroup
 }\\[1mm]
\scalebox{0.7}{% GNUPLOT: LaTeX picture with Postscript
\begingroup%
  \makeatletter%
  \newcommand{\GNUPLOTspecial}{%
    \@sanitize\catcode`\%=14\relax\special}%
  \setlength{\unitlength}{0.1bp}%
\begin{picture}(3600,2160)(0,0)%
\special{psfile=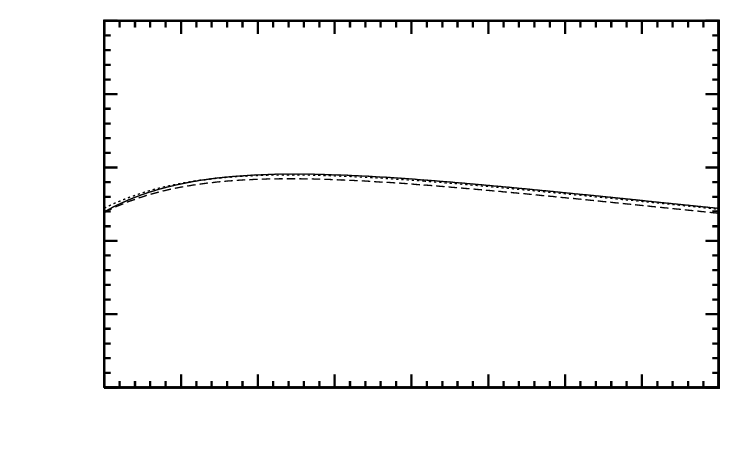 llx=0 lly=0 urx=360 ury=216 rwi=3600}
\put(1975,50){\makebox(0,0){\scalebox{1.3}{$\mu_c$ [GeV]}}}%
\put(100,1180){%
\special{ps: gsave currentpoint currentpoint translate
270 rotate neg exch neg exch translate}%
\makebox(0,0)[b]{\shortstack{\scalebox{1.3}{$P_c$}}}%
\special{ps: currentpoint grestore moveto}%
}%
\put(3450,200){\makebox(0,0){ 3}}%
\put(3081,200){\makebox(0,0){ 2.75}}%
\put(2713,200){\makebox(0,0){ 2.5}}%
\put(2344,200){\makebox(0,0){ 2.25}}%
\put(1975,200){\makebox(0,0){ 2}}%
\put(1606,200){\makebox(0,0){ 1.75}}%
\put(1238,200){\makebox(0,0){ 1.5}}%
\put(869,200){\makebox(0,0){ 1.25}}%
\put(500,200){\makebox(0,0){ 1}}%
\put(450,2060){\makebox(0,0)[r]{ 0.42}}%
\put(450,1708){\makebox(0,0)[r]{ 0.4}}%
\put(450,1356){\makebox(0,0)[r]{ 0.38}}%
\put(450,1004){\makebox(0,0)[r]{ 0.36}}%
\put(450,652){\makebox(0,0)[r]{ 0.34}}%
\put(450,300){\makebox(0,0)[r]{ 0.32}}%
\end{picture}%
\endgroup
 }
\end{center}
\vspace{-5mm}
\caption{$\Pc$ as a function of $\muc$ at NLO (upper plot) and
NNLO (lower plot). The three different lines correspond to three
different methods of computing $\as (\muc)$ from $\as (\MZ)$ (see
text).}        
\label{fig:plots}
\end{figure}

The dependence of $\Pc$ on $\muc$ can be seen in \Fig{fig:plots}. The 
solid line in the upper plot shows the NLO result obtained by 
evaluating $\as (\muc)$ from $\as (\MZ)$ solving the RG equation of
$\as$ numerically, while the dashed and dotted lines are obtained by
first determining the scale parameter $\Lambda_{\MSbar}$ from $\as
(\MZ)$, either using the explicit solution of the RG equation of $\as$
or by solving the RG equation of $\as$ iteratively for
$\Lambda_{\MSbar}$, and subsequently calculating $\as (\muc)$ from 
$\Lambda_{\MSbar}$. The corresponding two-loop values for $\as (\muc)$
have been obtained with the program {\tt RUNDEC}
\cite{Chetyrkin:2000yt}. Obviously, the difference between the 
three curves is due to higher order terms and has to be regarded as
part of the theoretical error. With its size of $\pm 0.012$ it is
comparable to the variation of the NLO result due to $\muc$, amounting
to $\pm 0.020$. In \cite{Buras:2004uu, Buchalla:1993wq,
Buchalla:1998ba} larger values for the latter uncertainty have been
quoted. The observed difference is related to the definition of the
charm quark mass. Replacing $\mc (\mc)$ in the logarithms $\ln (
\muc^2/\mc^2 )$ of the one-loop matrix elements by the more
appropriate $\mc (\muc)$ leads to a significant reduction of the
dependence of $\Pc$ on $\muc$. A detailed discussion of this issue
will be presented in \cite{comingsoon}. Finally, while in
\cite{Buras:2004uu, Buchalla:1993wq, Buchalla:1998ba} only $\muc$ was
varied, the theoretical error given in \Eq{eq:PcNLO} includes also the
dependence on $\mub$ and $\muh$ of $\pm 0.004$ and $\pm 0.001$,
respectively. The specified scale uncertainties correspond to the
ranges $1 \GeV \le \muc \le 3 \GeV$, $2.5 \GeV \le \mub \le 10 \GeV$, 
and $40 \GeV \le \muh \le 160 \GeV$.   

Using the input parameters listed in \Tab{tab:input}, we find from 
\Eqsto{eq:BR}{eq:PcNLO} at the NLO \footnote{Using the input
parameters of the UT{\it fit} Collaboration that are available at
http://utfit.roma1.infn.it/, we find $\BR (\Ktopinunu) = \left ( 7.86
\pm 0.76_{\Pc} \pm 0.72_{\rm other} \right ) \times 10^{-11}$ and $\BR
(\Ktopinunu) = \left ( 7.89 \pm 0.48_{\Pc} \pm 0.72_{\rm other} \right
) \times 10^{-11}$ at the NLO and the NNLO, respectively.}      
\begin{align} \label{eq:BRNLO}
\BR (\Ktopinunu) & = \left ( 7.93 \pm 0.77_{\Pc} \pm 0.84_{\rm other}
\right ) \times 10^{-11} \non \\
& = \left ( 7.9 \pm 1.3 \right ) \times 10^{-11} \, ,   
\end{align}
where the second error in the first line collects the uncertainties
due to $\kappa_+$, $\delta P_{c, u}$, $X (x_t)$, and the CKM
elements. The final error has only an illustrative character, as the
individual uncertainties have no statistical
interpretation. Numerically, the enhancement of $\BR (\Ktopinunu)$
coming from $\dPuc$ \cite{Isidori:2005xm} has been compensated by the 
suppression due to the decrease of $\mtpole$ \cite{Group:2005cc}.    

Provided $\Pc$ is known with a sufficient precision, a measurement
of $\Ktopinunu$, either alone or together with one of $\KLtopinunu$,
allows for precise determinations of the CKM parameters \cite{UT}. The
comparison of this unitarity triangle with the one from $B$ physics
offers a stringent and unique test of the SM. In particular for $\BR
(\Ktopinunu)$ and $\BR (\KLtopinunu)$ close to their SM predictions,
one finds that a given uncertainty $\sigma(\Pc)$ translates into
\begin{align} \label{eq:Vtderror}
\f{\sigma \left ( | V_{td} | \right)}{| V_{td} |} & = \pm 0.41 \, 
\f{\sigma \left (\Pc \right )}{\Pc} \, , \\ \label{eq:sinerror}
\f{\sigma \left ( \sin 2 \beta \right)}{\sin 2 \beta} & = \pm 0.34 \, 
\f{\sigma \left (\Pc \right )}{\Pc} \, , \\ \label{eq:gammaerror}
\f{\sigma \left ( \gamma \right)}{\gamma} & = \pm 0.83 \, \f{\sigma
\left (\Pc \right )}{\Pc} \, , 
\end{align}
with similar formulas given in \cite{Buras:2004uu}. Here $V_{td}$ is
the element of the CKM matrix and $\beta$ and $\gamma$ are the angles
in the standard unitarity triangle. As the uncertainties in
\Eqsand{eq:PcNLO}{eq:BRNLO} coming from the charm quark mass and the 
CKM parameters should be decreased in the coming years it is also
desirable to reduce the theoretical uncertainty in $\Pc$. To this end,
we here extend the NLO analysis of $\Pc$ presented in
\cite{Buchalla:1993wq, Buchalla:1998ba} to the next-to-next-to-leading
order (NNLO). This requires the computation of three-loop anomalous
dimensions of certain operators and of certain two-loop contributions. 

\begin{figure}[!t]
\begin{center}
\scalebox{0.5}{\includegraphics{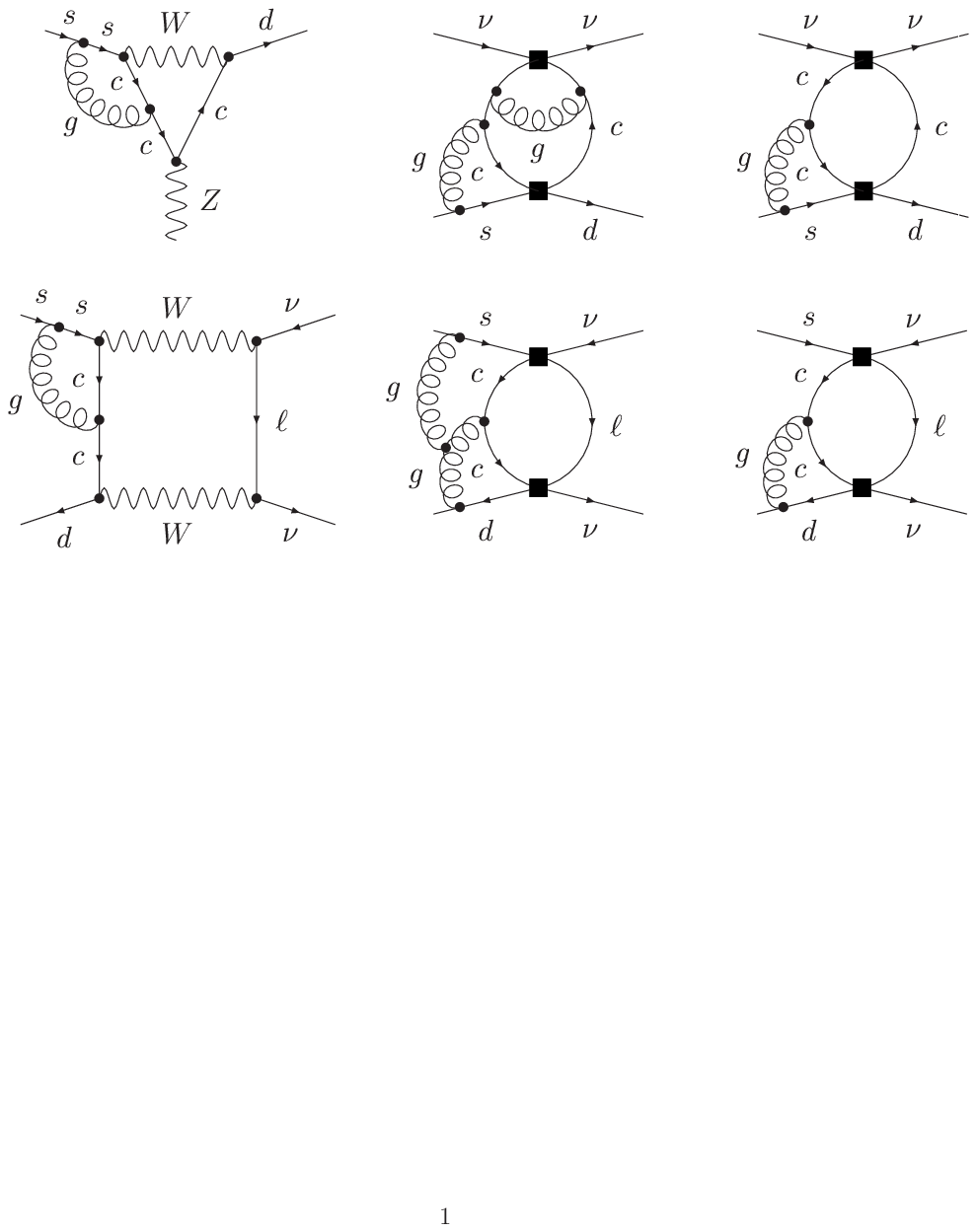}}
\end{center}
\vspace{-5mm}
\caption{Examples of Feynman diagrams arising in the full SM (left
column), describing the mixing of operators (center column) and the
matrix elements (right column) in the $Z$-penguin (upper row) and the
electroweak box (lower row) sector. Only the divergent pieces of the
diagrams displayed in the center column have to be computed, while the
Feynman graphs shown on the left- and right-hand side are needed
including their finite parts.}        
\label{fig:diagrams}
\end{figure}

The main components of the NNLO calculation \cite{comingsoon}, which
aims at resumming all $\ord (\as^n \ln^{n-1} ( \muh^2/\muc^2 ))$
logarithms in $\Pc$, are $(i)$ the $\ord (\as^2)$ matching corrections
to the relevant Wilson coefficients arising at $\muh$, $(ii)$ the $\ord
(\as^3)$ anomalous dimensions describing the mixing of the
dimension-six and -eight operators, $(iii)$ the $\ord (\as^2)$
threshold corrections to the Wilson coefficients originating at
$\mub$, and $(iv)$ the $\ord (\as^2)$ matrix elements of some of the
operators emerging at $\muc$. 

Conceptual new features in our NNLO calculation are $(a)$ the
appearance of the vector component of the effective neutral-current
coupling describing the interaction of neutrinos and quarks mediated
by $Z$-boson exchange, $(b)$ the presence of anomalous triangle
contributions which make it necessary to introduce a Chern-Simons
operator in order to obtain the correct anomalous Ward identity
involving the axial-vector coupling of the $Z$ boson, and $(c)$ the
existence of nontrivial two-loop matching corrections to the Wilson
coefficients of the current-current operators at the bottom quark
threshold.       

To determine the contributions of type $(i)$, $(iii)$, and $(iv)$ one
must calculate two-loop Green functions in the full SM and in
effective theories with five or four flavors. Sample diagrams for
steps $(i)$ and $(iv)$ are shown in the left and right columns of
\Fig{fig:diagrams}. The contributions $(ii)$ are found by calculating
three-loop Green functions with operator insertions. Sample diagrams
with a double insertion of dimension-six operators are shown in the
center column of \Fig{fig:diagrams}. The corresponding three-loop
amplitudes are evaluated using the method that has been described in
\cite{ADM, Gorbahn:2004my}. A comprehensive discussion of the
technical details of the matching, the renormalization of the 
effective theory and the actual calculation will be given in
\cite{comingsoon}.     

Having described the general steps of our calculation, we now present
our results. Using the general RG formalism \cite{Gorbahn:2004my,
NNLORGE}, we find at the NNLO  
\beq \label{eq:PcNNLO}
\begin{split}
\Pc & = 0.371 \pm 0.009_{\rm theor} \pm 0.031_{\mc} \pm 0.009_{\as}
\\ 
& = \left ( 0.37 \pm 0.04 \right ) \left ( \f{0.2248}{\lambda} \right
)^4 \, ,     
\end{split}
\eeq 
where the final error is fully dominated by the uncertainty in $\mc
(\mc)$. Comparing these numbers with \Eq{eq:PcNLO} we observe that our
NNLO calculation reduces the theoretical uncertainty by a factor of 4.  

As can be nicely seen in the lower plot of \Fig{fig:plots}, $\Pc$
depends very weakly on $\muc$ at the NNLO, varying by only $\pm
0.0047$. Furthermore, the three different treatments of $\as$ affect
the NNLO result by as little as $\pm 0.0005$. The three-loop values of
$\as (\muc)$ used in the numerical analysis have been obtained with
the program {\tt RUNDEC} \cite{Chetyrkin:2000yt}. The theoretical
error quoted in \Eq{eq:PcNNLO} includes also the dependence on $\mub$
and $\muh$ of $\pm 0.0028$ and $\pm 0.0007$, respectively. The
presented scale uncertainties correspond to the ranges given earlier.

Using \Eqs{eq:BR}, \eq{eq:kappap}, and \eq{eq:PcNNLO} the result in
\Eq{eq:BRNLO} is modified to the NNLO value 
\footnotemark[\value{footnote}] 
\begin{align} \label{eq:BRNNLO}
\BR (\Ktopinunu) & = \left ( 7.96 \pm 0.49_{\Pc} \pm 0.84_{\rm other}
\right ) \times 10^{-11} \non \\
& = \left ( 8.0 \pm 1.1 \right ) \times 10^{-11} \, . 
\end{align}

Employing \Eqsto{eq:Vtderror}{eq:gammaerror} the reduction of the
theoretical error in $\Pc$ from $\pm 10.1 \%$ down to $\pm 2.4 \%$
translates into the following uncertainties:     
\begin{align} \label{eq:errorcomparison}
\f{\sigma \left ( | V_{td} | \right)}{| V_{td} |} & = 
\begin{cases}
\pm 4.1 \% \, , & \hspace{1mm} \text{NLO} \, , \\
\pm 1.0 \% \, , & \hspace{1mm} \text{NNLO} \, , 
\end{cases} \\
\sigma \left ( \sin 2 \beta \right ) & = 
\begin{cases}
\pm 0.025 \, , & \text{NLO} \, , \\
\pm 0.006 \, , & \text{NNLO} \, , 
\end{cases} \\
\sigma \left ( \gamma \right) & = 
\begin{cases}
\pm 4.9^\circ \, , & \hspace{2mm} \text{NLO} \, , \\
\pm 1.2^\circ \, , & \hspace{2mm} \text{NNLO} \, , 
\end{cases} 
\end{align}
implying a very significant improvement of the NNLO over the NLO
results. In obtaining these numbers we have used $\sin 2 \beta =
0.724$ and $\gamma = 58.6^\circ$ \cite{Charles:2004jd}, and included
only the theoretical errors quoted in \Eqsand{eq:PcNLO}{eq:PcNNLO}. 

On the experimental side the Alternating Gradient Synchrotron E787 and
E949 Collaborations at Brookhaven observed the decay $\Ktopinunu$
finding three events so far \cite{experiment}. The resulting branching
ratio is 
\beq \label{eq:BRexp}
\BR (\Ktopinunu) = \left ( 14.7^{+13.0}_{-8.9} \right ) \times 10^{-11}
\, . 
\eeq 
Within theoretical, parametric, and experimental uncertainties,
\Eq{eq:BRNNLO} is fully consistent with the data. The prospects 
for the future measurements of $\BR (\Ktopinunu)$ can be found in 
\cite{prospects}.    
 
To conclude, we have evaluated the complete NNLO correction of the
charm quark contribution to $\BR (\Ktopinunu)$. The inclusion of these
contributions leads to a drastic reduction of the theoretical
uncertainty in the relevant parameter $\Pc$. This strengthens the
power of the $\Ktopinunus$ system in determining the CKM parameters
and increases its reach to new physics, in particular, if future
experimental values of $\BR (\Ktopinunu)$ will not differ much from
the SM prediction.    
 
We would like to thank G.~Buchalla, A.~H\"ocker, M.~Pierini,
F.~Schwab, and L.~Silvestrini for discussions. A.~J.~B.\ is supported
by the BMBF under contract 05HT4WOA/3. U.~H.\ and U.~N.\ are supported
by the DOE under contract DE-AC02-76CH03000.   \\

{\bf Note added:} Contrary to the statement made in point~$(b)$ in the second paragraph
after Eq.~(7) the initial condition of the  Chern-Simons operator vanishes. The numerical 
effect of this mistake is completely negligible, see the erratum of \cite{Buras:2006gb} for details.


\begin{thebibliography}{99}

\bibitem{UT}
G.~Buchalla and A.~J.~Buras,
%``Sin2beta from K $\to$ pi neutrino anti-neutrino,''
Phys.\ Lett.\ B {\bf 333}, 221 (1994);
%[arXiv:hep-ph/9405259] 
%%CITATION = HEP-PH 9405259;%%
%G.~Buchalla and A.~J.~Buras,
%``K $\to$ pi nu anti-nu and high precision determinations of  the CKM
%matrix,''
Phys.\ Rev.\ D {\bf 54}, 6782 (1996). 
%[arXiv:hep-ph/9607447].
%%CITATION = HEP-PH 9607447;%%

\bibitem{Gino}
G.~Isidori,
%``Kaon decays and the flavour problem,''
Annales Henri Poincare {\bf 4}, S97 (2003);
%[arXiv:hep-ph/0301159].
%%CITATION = HEP-PH 0301159;%%
%G.~Isidori,
%``$\Ktopinunus$ decays and CKM fits''
in Proceedings of the 2nd Workshop on the CKM Unitarity Triangle,
Durham, England, 2003, eConf {\bf C0304052}, WG304 (2003) and
references therein.  
%[arXiv:hep-ph/0307014].
%%CITATION = HEP-PH 0307014;%%

%%CITATION = HEP-PH 0301159;%%

\bibitem{Buras:2004uu}
A.~J.~Buras, F.~Schwab and S.~Uhlig,
%``Waiting for precise measurements of K+ $\to$ pi+ nu anti-nu and
%K(L) $\to$ pi0 nu anti-nu,''
hep-ph/0405132. 
%%CITATION = HEP-PH 0405132;%%

\bibitem{D'Ambrosio:2002ex}
G.~D'Ambrosio, G.~F.~Giudice, G.~Isidori and A.~Strumia,
%``Minimal flavour violation: An effective field theory approach,''
Nucl.\ Phys.\ B {\bf 645}, 155 (2002).
%[arXiv:hep-ph/0207036].
%%CITATION = HEP-PH 0207036;%%

\bibitem{Marciano:1996wy}
W.~J.~Marciano and Z.~Parsa,
%``Rare kaon decays with 'missing energy',''
Phys.\ Rev.\ D {\bf 53}, R1 (1996).
%%CITATION = PHRVA,D53,1;%%

\bibitem{Isidori:2005xm}
G.~Isidori, F.~Mescia and C.~Smith,
%``Light-quark loops in K $\to$ pi nu nu,''
Nucl.\ Phys.\ B {\bf 718}, 319 (2005).
%[arXiv:hep-ph/0503107].
%%CITATION = HEP-PH 0503107;%%

\bibitem{Isidori:2005tv}
G.~Isidori, G.~Martinelli and P.~Turchetti,
%``Rare kaon decays on the lattice,''
hep-lat/0506026.
%%CITATION = HEP-LAT 0506026;%%

\bibitem{Buchalla:1993wq}
G.~Buchalla and A.~J.~Buras,
%``The rare decays K+ $\to$ pi+ neutrino anti-neutrino and K(L) $\to$
%mu+ mu- beyond leading logarithms,''
Nucl.\ Phys.\ B {\bf 412}, 106 (1994).
%[arXiv:hep-ph/9308272].
%%CITATION = HEP-PH 9308272;%%

\bibitem{Buchalla:1998ba}
G.~Buchalla and A.~J.~Buras,
%``The rare decays K $\to$ pi nu anti-nu, B $\to$ X nu anti-nu and  B
%$\to$ l+ l-: An update,''
Nucl.\ Phys.\ B {\bf 548}, 309 (1999).
%[arXiv:hep-ph/9901288].
%%CITATION = HEP-PH 9901288;%%

\bibitem{Charles:2004jd}
J.~Charles {\it et al.}  [CKMfitter Group],
%``CP violation and the CKM matrix: Assessing the impact of the asymmetric B
%factories,''
Eur.\ Phys.\ J.\ C {\bf 41}, 1 (2005),
%[arXiv:hep-ph/0406184] 
and August 1, 2005 update available at
http://www.slac.stanford.edu/xorg/ckmfitter/ 
ckm\_results\_summerEPS2005.html.
%%CITATION = HEP-PH 0406184;%%

\bibitem{Bobeth:2003at}
C.~Bobeth, P.~Gambino, M.~Gorbahn and U.~Haisch,
%``Complete NNLO QCD analysis of anti-B $\to$ X/s l+ l- and higher order
%electroweak effects,''
JHEP {\bf 0404}, 071 (2004).
%[arXiv:hep-ph/0312090].
%%CITATION = HEP-PH 0312090;%%

\bibitem{Buchalla:1997kz}
G.~Buchalla and A.~J.~Buras,
%``Two-loop large-m(t) electroweak corrections to K $\to$ pi nu
%anti-nu for arbitrary Higgs boson mass,''
Phys.\ Rev.\ D {\bf 57}, 216 (1998).
%[arXiv:hep-ph/9707243].
%%CITATION = HEP-PH 9707243;%%

\bibitem{PDG}
S.~Eidelman {\it et al.}  [Particle Data Group],
%``Review of particle physics,''
Phys.\ Lett.\ B {\bf 592} (2004) 1, and 2005 partial update for
edition 2006 available at http://pdg.lbl.gov/.  
%%CITATION = PHLTA,B592,1;%%

\bibitem{X}
G.~Buchalla and A.~J.~Buras,
%``QCD corrections to the anti-s d Z vertex for arbitrary top quark mass,''
Nucl.\ Phys.\ B {\bf 398}, 285 (1993); 
%%CITATION = NUPHA,B398,285;%%
%G.~Buchalla and A.~J.~Buras  
%``QCD corrections to rare K and B decays for arbitrary top quark mass,''
%Nucl.\ Phys.\ B {\bf 400}, 225 (1993);
{\bf 400}, 225 (1993);
%%CITATION = NUPHA,B400,225;%%
M.~Misiak and J.~Urban,
%``{QCD} corrections to FCNC decays mediated by Z-penguins and W-boxes,''
Phys.\ Lett.\ B {\bf 451}, 161 (1999).
%[arXiv:hep-ph/9901278].
%%CITATION = HEP-PH 9901278;%%

\bibitem{Group:2005cc}
J.~F.~Arguin {\it et al.} [The Tevatron Electroweak Working Group],
%``Combination of CDF and D0 Results on the Top-Quark Mass,''
hep-ex/0507091.
%%CITATION = HEP-EX 0507091;%%

\bibitem{Melnikov:2000qh}
K.~Melnikov and T.~v.~Ritbergen,
%``The three-loop relation between the MS-bar and the pole quark masses,''
Phys.\ Lett.\ B {\bf 482}, 99 (2000).
%[arXiv:hep-ph/9912391].
%%CITATION = HEP-PH 9912391;%%

\bibitem{LO}
A.~I.~Vainshtein, V.~I.~Zakharov, V.~A.~Novikov and M.~A.~Shifman,
%``On The Strong Interaction Effects On The K(L) $\to$ 2 Mu Decay And
%K(L) K(S) Mass Difference. A Reply,''
Phys.\ Rev.\ D {\bf 16}, 223 (1977);
%%CITATION = PHRVA,D16,223;%%
J.~R.~Ellis and J.~S.~Hagelin,
%``Constraints On Light Particles From Kaon Decays,''
Nucl.\ Phys.\ B {\bf 217}, 189 (1983);
%%CITATION = NUPHA,B217,189;%%
C.~Dib, I.~Dunietz and F.~J.~Gilman,
%``Strong Interaction Corrections To The Decay K $\to$ Pi Neutrino
%Anti-Neutrino For Large M(T),''
Mod.\ Phys.\ Lett.\ A {\bf 6}, 3573 (1991).
%%CITATION = MPLAE,A6,3573;%%

\bibitem{charm}
J.~H.~K\"uhn and M.~Steinhauser,
%``Determination of alpha(s) and heavy quark masses from recent
%measurements of R(s),''
Nucl.\ Phys.\ B {\bf 619}, 588 (2001); {\bf 640}, 415(E) (2002);  
%[arXiv:hep-ph/0109084];
%%CITATION = HEP-PH 0109084;%%
J.~Rolf and S.~Sint  [ALPHA Collaboration],
%``A precise determination of the charm quark's mass in quenched QCD,''
JHEP {\bf 12}, 007 (2002);
%[arXiv:hep-ph/0209255];
%%CITATION = HEP-PH 0209255;%%
A.~H.~Hoang and M.~Jamin,
%``MS-bar charm mass from charmonium sum rules with contour
%improvement,'' 
Phys.\ Lett.\ B {\bf 594}, 127 (2004).
%[arXiv:hep-ph/0403083].
%%CITATION = HEP-PH 0403083;%%

\bibitem{Chetyrkin:2000yt}
K.~G.~Chetyrkin, J.~H.~K\"uhn and M.~Steinhauser,
%``RunDec: A Mathematica package for running and decoupling of the
%strong coupling and quark masses,''
Comput.\ Phys.\ Commun.\  {\bf 133}, 43 (2000).
%[arXiv:hep-ph/0004189].
%%CITATION = HEP-PH 0004189;%%

\bibitem{comingsoon}
A.~J.~Buras, M.~Gorbahn, U.~Haisch and U.~Nierste, in preparation. 

\bibitem{ADM}
M.~Misiak and M.~M\"unz,
%``Two loop mixing of dimension five flavor changing operators,''
Phys.\ Lett.\ B {\bf 344}, 308 (1995);
%[arXiv:hep-ph/9409454];
%%CITATION = HEP-PH 9409454;%%
K.~G.~Chetyrkin, M.~Misiak and M.~M\"unz,
%``Beta functions and anomalous dimensions up to three loops,''
Nucl.\ Phys.\ B {\bf 518}, 473 (1998);
%[arXiv:hep-ph/9711266].
%%CITATION = HEP-PH 9711266;%%
P.~Gambino, M.~Gorbahn and U.~Haisch,
%``Anomalous dimension matrix for radiative and rare semileptonic B
%decays up to three loops,''
Nucl.\ Phys.\ B {\bf 673}, 238 (2003);
%[arXiv:hep-ph/0306079];
%%CITATION = HEP-PH 0306079;%%
M.~Gorbahn, U.~Haisch and M.~Misiak,
%``Three-loop mixing of dipole operators,''
hep-ph/0504194.
%%CITATION = HEP-PH 0504194;%%

\bibitem{Gorbahn:2004my}
M.~Gorbahn and U.~Haisch,
%``Effective Hamiltonian for non-leptonic $|$Delta(F)$|$ = 1 decays
%at NNLO in QCD,''
Nucl.\ Phys.\ B {\bf 713}, 291 (2005). 
%[arXiv:hep-ph/0411071].
%%CITATION = HEP-PH 0411071;%%

\bibitem{NNLORGE}
M.~Beneke, T.~Feldmann and D.~Seidel,
%``Systematic approach to exclusive B $\to$ V l+ l-, V gamma decays,''
Nucl.\ Phys.\ B {\bf 612}, 25 (2001); 
%[arXiv:hep-ph/0106067]; 
%%CITATION = HEP-PH 0106067;%%
H.~M.~Asatrian, K.~Bieri, C.~Greub and M.~Walker, 
%``Virtual- and bremsstrahlung corrections to b $\to$ d l+ l- in the
%standard model,''
Phys.\ Rev.\ D {\bf 69}, 074007 (2004). 
%[arXiv:hep-ph/0312063]. 
%%CITATION = HEP-PH 0312063;%%

\bibitem{experiment}
S.~C.~Adler {\it et al.}  [E787 Collaboration],
%``Evidence for the decay K+ $\to$ pi+ nu anti-nu,''
Phys.\ Rev.\ Lett.\  {\bf 79}, 2204 (1997);
%[arXiv:hep-ex/9708031],  
%%CITATION = HEP-EX 9708031;%%
%S.~C.~Adler {\it et al.}  [E787 Collaboration],
%``Further search for the decay $K^+ \to  \pi^+ \nu \bar{\nu},''
%Phys.\ Rev.\ Lett.\  {\bf 84}, 3768 (2000);
{\bf 84}, 3768 (2000);
%[arXiv:hep-ex/0002015], 
%%CITATION = HEP-EX 0002015;%%
% S.~Adler {\it et al.}  [E787 Collaboration],
%``Further evidence for the decay K+ $\to$ pi+ nu anti-nu,''
%Phys.\ Rev.\ Lett.\  {\bf 88}, 041803 (2002)
{\bf 88}, 041803 (2002); 
%[arXiv:hep-ex/0111091] 
%%CITATION = HEP-EX 0111091;%%
%S.~Adler {\it et al.}  [E787 Collaboration],
%``Further search for the decay K+ $\to$ pi+ nu anti-nu in the
%momentum region P < 195-MeV/c,''
Phys.\ Rev.\ D {\bf 70}, 037102 (2004);
%[arXiv:hep-ex/0403034]; 
%%CITATION = HEP-EX 0403034;%%
V.~V.~Anisimovsky {\it et al.}  [E949 Collaboration],
%``Further study of the decay K+ $\to$ pi+ nu anti-nu,''
Phys.\ Rev.\ Lett.\  {\bf 93}, 031801 (2004).
%[arXiv:hep-ex/0403036].
%%CITATION = HEP-EX 0403036;%%
 
\bibitem{prospects}
D.~Bryman,
%``Rare kaon decays: Progress and prospects,''
hep-ex/0206072;
%%CITATION = HEP-EX 0206072;%%
L.~Littenberg,
%``Rare kaon and pion decays,''
hep-ex/0212005;
%%CITATION = HEP-EX 0212005;%%
NA48 Collaboration, (2005), http://
na48.web.cern.ch/NA48/NA48-3/. 

\bibitem{Buras:2006gb} 
  A.~J.~Buras, M.~Gorbahn, U.~Haisch and U.~Nierste,
  %``Charm quark contribution to K+ ---> pi+ nu anti-nu at next-to-next-to-leading order,''
  JHEP {\bf 0611}, 002 (2006)
  [hep-ph/0603079v3].
  %%CITATION = HEP-PH/0603079;%%

\end{thebibliography}
\end{document}